\documentclass{article}
\usepackage{spconf,amsmath,graphicx,hyperref}
\usepackage{booktabs}

\title{StreamMark: A Deep Learning-Based Semi-Fragile Audio Watermarking for Proactive Deepfake Detection}
\twoauthors
 {Zhentao Liu\sthanks{The first author performed the work during the internship at Logitech}}
	{EPFL, Switzerland}
 {Milos Cernak}
	{Logitech Europe, Switzerland}
\begin{document}
\ninept
\maketitle
\begin{abstract}
The rapid advancement of generative AI has made it increasingly challenging to distinguish between deepfake audio and authentic human speech. To overcome the limitations of passive detection methods, we propose StreamMark, a novel deep learning-based, semi-fragile audio watermarking system. StreamMark is designed to be robust against benign audio conversions that preserve semantic meaning (e.g., compression, noise) while remaining fragile to malicious, semantics-altering manipulations (e.g., voice conversion, speech editing). Our method introduces a complex-domain embedding technique within a unique Encoder-Distortion-Decoder architecture, trained explicitly to differentiate between these two classes of transformations. Comprehensive benchmarks demonstrate that StreamMark achieves high imperceptibility (SNR 24.16 dB, PESQ 4.20), is resilient to real-world distortions like Opus encoding, and exhibits principled fragility against a suite of deepfake attacks, with message recovery accuracy dropping to chance levels (\texttt{-}50\%), while remaining robust to benign AI-based style transfers (ACC $>98\%$). 
\end{abstract}

\section{Introduction}
The escalating sophistication of generative speech models presents a significant threat to the integrity of digital communication. Technologies such as neural voice cloning and zero-shot text-to-speech (TTS) can now synthesize voices that are virtually indistinguishable from those of real individuals~\cite{wang2023neural,casanova2022yourtts}, creating significant security and authenticity risks, particularly in enterprise environments where trusted communication is paramount.

Historically, the primary defense against such synthetic media has been passive detection~\cite{lv2022fake,wang2020deepsonar}, which typically involves training Machine Learning (ML) classifiers to identify artifacts unique to generated content. However, this approach is fundamentally reactive and suffers from several critical drawbacks. As generative models continuously improve, the subtle differences between authentic and synthetic audio diminish, leading to a continuous challenge, where detectors quickly become obsolete~\cite{san2024proactive}. These passive methods also exhibit limited generalization to new or unseen synthesis techniques and are demonstrably vulnerable to adversarial attacks designed to evade detection. Furthermore, the very definition of "fake" audio becomes ambiguous in real-world applications; for instance, it is unclear whether audio that has been legitimately enhanced with AI-based denoising should be flagged as inauthentic, a nuance that simple binary classifiers struggle to handle.

To circumvent these limitations, digital watermarking offers a proactive defense mechanism. By embedding an imperceptible, verifiable signal directly into an audio stream at its source, watermarking can establish a definitive chain of provenance and integrity. The growing importance of this approach is underscored by recent regulatory initiatives in the United States, the European Union, and China, which are moving towards mandating the watermarking of AI-generated content to ensure transparency and accountability.

Despite its promise, the field of audio watermarking has traditionally focused on the goal of robustness. Both classic Digital Signal Processing (DSP) methods \cite{Patchwork, 8682352, 923725} and modern Deep-Learning-based Audio Watermarking (DLAW) frameworks \cite{san2024proactive, liu2023detecting, MaskMark}, have been engineered to ensure that an embedded watermark survives any and all signal transformations. This focus on robustness, however, represents a conceptual flaw when applied to the problem of deepfake authentication. A watermark that successfully survives a malicious manipulation, such as the complete replacement of a speaker's voice, fails in its primary purpose—to signal that the audio's semantic content has been compromised. The very property of robustness becomes a liability.

This paper introduces a fundamental shift in this paradigm. We propose StreamMark, the first deep-learning-based semi-fragile audio watermarking framework designed specifically for deepfake detection. The concept of semi-fragility, which we adapt from the domain of image forensics~\cite{neekhara2024facesigns,beuve2023waterlo}, redefines the objective of watermarking. A semi-fragile watermark is engineered to be robust against benign, semantics-preserving conversions (e.g., noise, compression, style transfer ~\cite{steinmetz2022style}) but fragile against malicious, semantics-altering conversions (e.g., Text-to-Speech (TTS) ~\cite{zhang2023speak}, Voice Conversion (VC) ~\cite{li2023freevc}, and Speech Editing ~\cite{peng2024voicecraft}). This dual behavior allows the watermark to function as an indicator of semantic integrity. Our primary contributions include the novel StreamMark architecture, which features a complex-domain embedding technique for enhanced imperceptibility and a unique training objective that explicitly teaches the model to differentiate between benign and malicious transformations. We validate StreamMark through a comprehensive set of classic and deepfake-centric benchmarks, demonstrating its superior performance and its practical applicability in real-time enterprise headset scenarios.

\section{Related work}
\label{sec:RelatedWork}



The foundations of audio watermarking lie in DSP techniques, which are broadly categorized into time-domain \cite{Patchwork,1580525} and transform-domain \cite{923725, 6200333} methods. While foundational, these hand-crafted methods rely on specific signal properties and struggle to withstand the complex, non-linear distortions introduced by modern AI-based attacks and advanced audio codecs, motivating the transition to data-driven, deep learning approaches.

Since 2022, DLAW has emerged as the state-of-the-art, leveraging neural networks to learn optimal embedding strategies automatically. The field began with Pavlovic's encoder-detector architecture~\cite{PAVLOVIC2022103381}, which demonstrated the potential of deep learning for robust watermarking. Subsequent works have refined this concept with more advanced architectures and objectives. MaskMark~\cite{MaskMark} introduced a multiplicative spectrogram mask to improve robustness over earlier additive methods. WavMark~\cite{chen2023wavmark} pioneered the use of invertible neural networks to better preserve signal quality and tackled the critical challenge of watermark localization within a longer audio stream, though it faced limitations in detection efficiency. Addressing this, AudioSeal~\cite{san2024proactive}, developed by Meta, proposed a highly efficient generator-detector architecture optimized for fast, localized watermark detection, making it suitable for large-scale, real-time applications.

Recent work has continued to focus on robustness in realistic deployment conditions ~\cite{singh24_interspeech, gan2025syncguardrobustaudiowatermarking}. Juvela et al. ~\cite{icassp/JuvelaW25} introduced codec-aware collaborative watermarking, where codec-like augmentations are integrated into training to ensure watermark robustness under lossy distribution pipelines. Xu et al. ~\cite{xu25f_interspeech} proposed WAKE, an INN-based approach with key enrichment, enabling secure extraction while maintaining audio quality. Özer et al. ~\cite{ozer25_interspeech}   released RAW-Bench, a standardized benchmark for evaluating watermark robustness under diverse real-world conditions, including neural codecs, reverberation, and compound transformations. Alongside these developments, recent evaluations  ~\cite{oreilly2025deep} highlight that many post-hoc watermarking schemes remain vulnerable to removal attacks, underscoring the need for designs that balance imperceptibility, robustness, and security.

\subsection{The Semi-Fragility Property}
\label{sec:semi-fragility-property}

A critical analysis of the DLAW landscape reveals that all major existing methods share a common, and for our purposes, limiting, objective: maximizing robustness. This is exemplified by Timbre Watermarking~\cite{liu2023detecting}, a method specifically designed to be robust against voice cloning attacks, treating them as a transformation the watermark must survive. While effective for its intended purpose of tracing a voice clone back to its source audio, this approach is fundamentally misaligned with the goal of authenticating the semantic integrity of an audio stream. If the watermark persists after a speaker's voice has been replaced, it fails to indicate that a malicious tampering event has occurred. The application of watermarking to deepfake detection necessitates a paradigm shift from ensuring robustness against signal degradation to enabling the detection of semantic manipulation.

This conceptual gap leads us to the domain of image forensics, where the idea of semi-fragile watermarking has been successfully implemented. Works such as FaceSigns~\cite{neekhara2024facesigns} and WaterLo~\cite{beuve2023waterlo} have developed deep learning frameworks that are robust to benign image processing (e.g., JPEG compression) but fragile to malicious manipulations (e.g., face swapping). They achieve this by training an encoder-decoder system with a specialized distortion layer that includes examples of both benign and malicious transformations, forcing the model to learn the distinction. StreamMark is the first work to adapt and extend this powerful semi-fragile concept to the audio domain, addressing a critical and previously unmet need in the fight against deepfake audio.

\begin{figure*}[t!]
    \centering
    \includegraphics[width=0.98\textwidth]{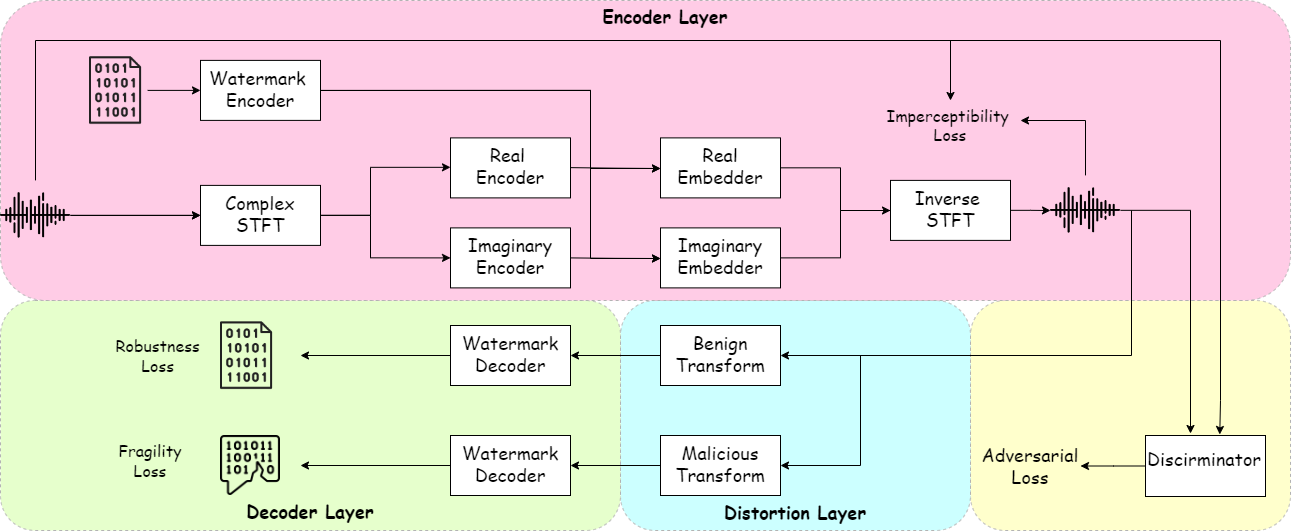}

   \caption{
      {StreamMark Model Architecture. The \textbf{Encoder Layer} is responsible for imperceptibly embedding a watermark message into a raw audio signal: (i) audio is first transformed to frequency domain yielding separate real and imaginary components, (ii) processed by parallel real and imaginary encoders to produce watermark carriers (iii) used to embed encoded message, and finally (iv) the watermarked real and imaginary components are transformed back into the time domain to produce the watermarked audio. The \textbf{Distortion Layer} contains two parallel sets of transformations that are randomly applied during training: (i) a benign conversion set $G_b$, and (ii) a malicious conversion set $G_m$. The \textbf{Decoder Layer} is tasked with extracting the watermark from a potentially distorted audio signal: (i) transforming the input audio to frequency domain, (ii) extracting a watermark features by processioning the real and imaginary components, and (iii) recovering the final message by the message decoder. 
      }}
    \label{fig:arch}
    
\end{figure*}

\section{The StreamMark method}
\label{sec:streammark}

The StreamMark framework formulates the thread model by considering attacker who performs benign conversions that destroy the watermark (a robustness attack) and to perform malicious conversions that preserve the watermark (an integrity attack).
To counter this threat, we formalize the concept of semi-fragility by defining two distinct classes of audio transformations:

\textbf{\textit{Benign Conversion}}: Any transformation performed within the standard audio pipeline that does not intentionally violate the audio's semantic meaning. This includes operations like adding noise, applying compression, or changing the audio style ~\cite{steinmetz2022style} (e.g., simulating a different microphone). The watermark must be recoverable after such transformations.
    
\textbf{\textit{Malicious Conversion}}: Any transformation performed outside the standard audio pipeline that intentionally alters the audio's core semantic integrity, such as the speaker's identity ("who") or the spoken content ("what"). This includes deepfake techniques like streaming Zero-Shot TTS, VC, and AI-based speech editing. StreamMark is designed to break (fragility) under such semantic manipulation. This ensures that any generated deepfake--even one derived from the watermarked source--fails verification.

\subsection{Network Architecture}

Fig.~\ref{fig:arch} shows
StreamMark that employs a three-layer architecture consisting of an Encoder, a Distortion Layer, and a Decoder, which are trained jointly in an end-to-end fashion. In the encoder layer, the watermark message is embedded in the STFT complex domain. In the distortion layer, benign transformations and malicious transformations are performed to
 simulate different audio conversions.

Real and imaginary Encoders, Embeddeders and Decoder are simple fully 2D convolutional networks with 6 layers, which employ a skip gated block as their fundamental unit. Watermak Encoder is a dense layer of size 512 with LeakyReLU activation function, whereas the output of the Watermark Decoder processed by a linear dense layer of the same size. The decoder utilizes an average pooling across the time dimension, a critical feature that provides robustness against desynchronization attacks like cropping and packet loss. 

A key innovation of StreamMark is its method of embedding the watermark in the complex domain of the STFT. Traditional DLAW methods, such as Timbre Watermarking, typically embed information only in the magnitude spectrogram, discarding the phase information or using it solely for reconstruction. However, psychoacoustic principles suggest that human hearing is less sensitive to phase distortions than to magnitude distortions in speech signals. By embedding perturbations in both the real and imaginary components, which correspond to both magnitude and phase, StreamMark can achieve a higher degree of imperceptibility. Our empirical investigations revealed that a naive attempt to embed the watermark only in the phase domain led to training instability and non-convergence. The complex-domain approach, where the network learns to optimally distribute the watermark across both real and imaginary parts, proved to be a more stable and effective solution for minimizing audible artifacts.

\subsection{Semi-Fragile Training Objective}
\label{sec:semi-fragile}

The central mechanism enabling StreamMark's semi-fragile behavior is its unique \textbf{Distortion Layer} and the associated training objective. Unlike prior DLAW frameworks that use a distortion layer with only benign transformations, StreamMark's layer contains two parallel transformation sets that are randomly applied during training:
\begin{enumerate}
    \item A benign conversion set ($G_b$), which includes operations like cropping, Gaussian noise, resampling, filtering, and requantization, which approximate non-adversarial signal-processing distortions that typically arise from standard recording, transmission, and storage procedures.
    \item A malicious conversion set ($G_m$), which simulates deepfake attacks. Deepfake attacks usually target the timbral characteristics of speech. To mimic this, we use pitch shifting to perform malicious conversions, effectively simulating the timbre changes in audio Deepfake.
\end{enumerate}

This dual-path distortion layer allows for the formulation of a composite loss function that explicitly optimizes for semi-fragility. The total loss $L$ is defined as: $L=\lambda_iL_i+\lambda_dL_d+\lambda_rL_r-\lambda_fL_f$.
Each component of the loss function targets a specific property:
\begin{itemize}
    \item $L_i$: An imperceptibility loss (MSE between original and watermarked audio) to ensure audio quality.
    \item $L_d$: An adversarial discriminator loss to further enhance imperceptibility by making watermarked audio indistinguishable from original audio.
    \item $L_r$: A robustness loss, defined as the MSE between the original message and the message recovered from an audio that has undergone a benign transformation from $G_b$. This term is minimized.
    \item $L_f$: A fragility loss, defined as the MSE between the original message and the message recovered from an audio that has undergone a malicious transformation from $G_m$. This term is maximized by applying a negative weight ($-\lambda_f$).
\end{itemize}

This objective function creates a minimax-like optimization problem. The network is trained to minimize the decoding error for benignly transformed audio while simultaneously maximizing the decoding error for maliciously transformed audio. This forces the encoder and decoder to learn a watermarking scheme that is inherently sensitive to the semantic nature of the transformations applied to the audio signal.

\section{Experimental evaluation}
\label{sec:experiments}
We selected two recent state-of-the-art DLAW techniques as the baselines: the \textbf{Timbre}~\cite{liu2023detecting} and Meta's \textbf{AudioSeal}~\cite{san2024proactive}. We included also the \textbf{Patchwork} system~\cite{Patchwork}, a classic DSP-based technique.

All DLAW techniques were trained on the same data, the \texttt{train\_clean100} subset of the Librispeech dataset~\cite{panayotov2015librispeech}, and evaluated on 500 randomly selected recordings from the \texttt{test\_clean} set that we post-processed and created two benchamrking datasets: a classic watermarking, the \textbf{test set A}, for imperceptibility and robustness testing, and the deepfake, \textbf{test set B}, for validating semi-fragility.

The test set A augmentation includes: audio cropping, MP3 encoding at various bit rates, and Opus encoding with different frame durations. Opus encoding is not used in the SteramMark's distortion layer, and thus it is an unknown attack.

The test set B is designed and open-sourced a as new Deepfake benchmark\footnote{\url{https://github.com/L1uZhentao/deepfake_benchmark}} to assess the semi-fragility property (introduced in Sec.~\ref{sec:semi-fragility-property}). Briefly, the benchmark is designed for assessing deepfakes, or to be more
general, under AI conversions. The AI conversion could be either malicious conversion or benign conversion. The boundary of benign/malicious AI conversion is whether the audio semantics is changed intentionally. The benchmark thus includes the following malicious conversions: (i) multilingual text-to-speech (TTS) conversions, incl. (ii) zero-shot multi-speaker TTS, (iii) Voice Conversion (VC), and (iv) Speech Editing conversions. The benchmark includes also a Style Transfer as a benign conversion.

The StreamMark model was trained with dynamic augmentation from the $G_b$ and $G_m$ sets as outlined in Sec.~\ref{sec:semi-fragile}. We used Adam optimizer ($\beta_1=0.94$, $\beta_2=0.98$), learning rate 0.0002, and the loss function $L$ with $\lambda_i=\lambda_d=0.01$ and $\lambda_r=\lambda_f=1.0$. We traind the model on two GPU NVIDIA GeForce RTX 2080 (8G). 
The number of encoders' parameters are as follows: StreamMark 0.9M, Timbre 0.45M, and AudioSeal 7.3M.


We used a fixed watermark message length of 16 bits for all evaluations. The primary evaluation metrics were Signal-to-Noise Ratio (SNR), Perceptual Evaluation of Speech Quality (PESQ) , and Speaker Encoder Cosine Similarity (SECS)  for imperceptibility, and Message Recovery Accuracy (ACC) for robustness and fragility.

\subsection{Imperceptibility and Robustness (Test Set A)}

The evaluation results, summarized in Tab.~\ref{tab:1}, demonstrate StreamMark's strong performance on classic watermarking benchmarks. In terms of imperceptibility, StreamMark achieves a PESQ score of 4.2, significantly outperforming the robust baseline Timbre Watermarking (3.7) and indicating high perceptual quality ($> 4.0$ MOS)  that is also achieved by the AudioSeal and the Patchwork baselines. This result validates the effectiveness of the complex-domain embedding strategy in minimizing audible artifacts.

StreamMark also exhibits exceptional robustness against a wide range of challenging, real-world benign conversions. It maintains near-perfect accuracy even under large-scale cropping (99.97\% ACC with 70\% of the audio removed), severe MP3 compression (87.26\% ACC at a low bitrate of 8 kbps), and, most notably, Opus encoding. The near-perfect accuracy ($>99.89$\%) against the Opus codec, which is prevalent in real-time communication platforms like Web-RTC, directly confirms StreamMark's suitability for its target application in enterprise headsets and online meetings.

\begin{table}[ht]
    \centering
    \caption{Imperceptibility and Robustness Evaluation. StreamMark demonstrates a superior balance of high perceptual quality, while keeping high message recovery accuracy (ACC) to challenging benign conversions compared to baselines of MP3 encoding at 8 kbps, and Opus with 60 ms frames. MP3 and Opus encodings are not used in the distortion layer, and thus they represent unknown attacks.}
    \label{tab:1}
    \resizebox{\linewidth}{!}{%
    \begin{tabular}{l|cccccc}
    \toprule
    Method & SNR & PESQ & SECS & Crop & MP3 & Opus \\
    & (dB) & & & (70\%) & (ACC) & (ACC)\\
    \midrule
    Patchwork & \textbf{33.65} & \textbf{4.34} & 0.99 & 0.72 & 0.61 & 0.85 \\
    AudioSeal & 25.41 & 4.30 & 0.99 & \textbf{1.00} & 0.85 & 0.57\\
    Timbre & 24.14 & 3.70 & 0.99 & 0.99 & 0.79 & 0.99\\
    \midrule
    StreamMark & 24.16 & 4.20 & 0.99 & 0.99 & \textbf{0.87} & \textbf{0.99} \\
    \bottomrule
    \end{tabular}
    }
\end{table}

\subsection{The Deepfake Benchmark (Test Set B)}

The core contribution of StreamMark is its semi-fragile nature, which was rigorously tested using a custom deepfake benchmark. This benchmark evaluates the model's response to both malicious, semantics-altering AI conversions and benign, semantics-preserving AI conversions. The results, presented in Tab.~\ref{tab:2}, provide clear evidence of StreamMark's principled semi-fragility.

When subjected to malicious deepfake attacks, the watermark's integrity is destroyed. For a range of state-of-the-art models performing Text-to-Speech~\cite{zhang2023speak}, Voice Conversion~\cite{li2023freevc}, and Speech Editing ~\cite{peng2024voicecraft}, the message recovery accuracy (ACC) drops to approximately 50\%, which is equivalent to random guessing for a binary message. This demonstrates that StreamMark successfully identifies these manipulations by becoming fragile, thereby signaling a breach of semantic integrity.

In stark contrast, when subjected to a benign AI-based Style Transfer~\cite{steinmetz2022style}, which alters the acoustic properties of the audio (e.g., simulating a different microphone) without changing the audio sematics (the speaker's identity or the content), the watermark remains highly robust. The ACC for various style transfers was consistently above 98\%. This crucial result confirms that StreamMark's fragility is not a naive reaction to any form of neural network processing but is specifically and correctly triggered by transformations that manipulate the fundamental semantics of the speech signal.

\begin{table}[ht]
    \centering
    \caption{The Deepfake Benchmark: Validating Semi-Fragility. StreamMark's message recovery accuracy (ACC) drops to 50\% for malicious conversions but remains high for benign AI-based style transfer, confirming its ability to detect semantic tampering.}
    \label{tab:2}
    \resizebox{\linewidth}{!}{%
    \begin{tabular}{l|ccc}
    \toprule
     Type& Model/Style & ACC (\%) & Exp. Behavior \\
    \midrule
    Malicious & VALL-E-X ~\cite{zhang2023speak} & 51.01 & Fragile \\
    (Fragility) & (TTS) & & (Destroyed)\\
    & FreeVC~\cite{li2023freevc} & 49.75 & Fragile\\
    &  (VC) & & (Destroyed)\\
    & VoiceCraft ~\cite{peng2024voicecraft} & 51.79 & Fragile\\
    &  (Editing) & & (Destroyed)\\
    \midrule
    Benign & DeepAFX ~\cite{steinmetz2022style} & 100.00 & Robust\\
    (Robustness) & (Bright) & & (Preserved)\\
    & DeepAFX & 98.73 & Robust\\
    &  (Broadcast) & & (Preserved)\\
    & DeepAFX  & 98.34 & Robust\\
    &  (Telephone) & & (Preserved)\\
    \bottomrule
    \end{tabular}
    }
\end{table}



\section{Conclusion}
\label{sec:conclusion}
This paper introduced StreamMark, a novel deep-learning-based semi-fragile audio watermarking framework designed as a proactive defense against the growing threat of deepfake audio. By adapting the semi-fragile paradigm from image processing to the audio domain, StreamMark addresses a critical limitation of existing watermarking methods, which have focused exclusively on robustness.

Our key findings demonstrate that StreamMark successfully balances the competing requirements of a modern authentication system. It achieves high imperceptibility, validated by objective metrics, and shows strong robustness against a wide array of benign digital conversions and real-world distortions, including the challenging Opus codec used in live communication or YouTube's encoding. Most importantly, our deepfake benchmark confirms its principled semi-fragile behavior: the watermark is reliably destroyed by malicious, semantics-altering manipulations such as voice conversion and speech editing, while remaining intact through benign, semantics-preserving AI transformations like style transfer.

The semi-fragile approach represents a significant advancement toward more reliable and trustworthy audio authentication in an era of ubiquitous generative AI. Our contribution is also open sourcing the Deepfake Benchamark used in this paper\footnote{\url{https://github.com/L1uZhentao/deepfake_benchmark}}.

Future work could extend the StreamMark framework to handle multi-channel audio streams, investigate its performance against more complex, composite deepfake attacks (that we aim to add to the benchmark), and further optimize the model's computational complexity for more efficient processing.

\vfill\pagebreak



\bibliographystyle{IEEEbib}
\bibliography{zliurefs,refs}

\end{document}